# Predicting Relevance Scores for Triples from Type-Like Relations using Neural Embedding

## The Cabbage Triple Scorer at WSDM Cup 2017

Yael Brumer, Bracha Shapira, Lior Rokach
Department of Software and Information Systems Engineering
and Telekom Innovation Laboratories
Ben-Gurion University of the Negev
{ybrumer,bshapira,liorrk}@bgu.ac.il

Oren Barkan
School of Computer Science
Tel Aviv University
orenbarkan@post.tau.ac.il

## ABSTRACT

The WSDM Cup 2017 Triple scoring challenge is aimed at calculating and assigning relevance scores for triples from type-like relations. Such scores are a fundamental ingredient for ranking results in entity search. In this paper, we propose a method that uses neural embedding techniques to accurately calculate an entity score for a triple based on its nearest neighbor. We strive to develop a new latent semantic model with a deep structure that captures the semantic and syntactic relations between words. Our method has been ranked among the top performers with accuracy - 0.74, average score difference - 1.74, and average Kendall's Tau - 0.35.

## 1. INTRODUCTION

Organizing the world's facts and storing them in a structured database, such as Freebase [3], have become important resources for supporting open-domain question answering (QA). In contrast to full-text search methods (e.g., Google), in which results are a mixture of relevant and irrelevant hits, knowledge based queries have a well-defined result set. Moreover, it is often desirable to rank the result sets, because users prefer to see the most relevant results first, regardless of the size of the returned set [1].

The annual WSDM Cup 2017 challenge is to compute relevance scores for triples from type-like relations, more specifically, for profession / nationality types [6]. The relevance score is used to signify "belonging" and quantifies the relationship between entities and types in terms of the degree to which an entity belongs to a type. Participants were provided several text files[1], including: 1) `persons` - contains more than 380,000 different entity names used for this task, 2) `profession.kb/nationality.kb` - each contains all of the profession / nationality tuples for a set of people, respectively, 3) `profession.train/nationality.train` - contains a subset of tuples from the respective .kb file together with their relevance scores generated based on human judgment, 4) `professions/nationalities` - contains a list of the 200 / 100 different professions / nationalities from the respective .kb file, and 5) `wiki-sentences` - contains more than 33 million sentences from Wikipedia with annotations of these people. Participants were free to use all of the data above, but could also consider any kind or amount of other data as well.

Inspired by the recent success of continuous space word representations in capturing the semantic similarities in various natural language processing tasks, we propose to incorporate neural embedding techniques to reveal semantic relations between entities and iteratively propagate their relevance scores through similar (according to word2vec vectors) neighbors. One can look at it as a kind of graph where each node represents an "entity" (from the `persons` file) that contains information about his/her relation types (profession or nationality). In this work, we introduce a neural model, capable of encoding large amounts of Wikipedia data (more than 33 million sentences), to learn representational embeddings for words. Each entity from `persons` consists of a large number of wiki-sentences (ranging from 68,662 sentences for the most frequently mentioned person, to three sentences for the least frequently mentioned person in this file).

The nearest neighbor approach is widely used today, integral to popular practices such as collaborative filtering in recommendation systems [9] and pattern classification [4]. While the underlying assumption guiding this work is that closer people (nearest neighbors) will share characteristics, our method was found to be more successful with regard to the profession relation (nearest neighbors have similar professions). We've made several adjustments to our approach in order to meet the needs of the nationality aspect; this work presents our comprehensive solution to the two-pronged task.

Our method has two advantages: 1) it facilitates the efficient and logical identification of new people and 2) because we propagate the knowledge with respect to the similarity score, the differences observed (person to person) may be more subtle, and thus the knowledge gained in each iteration will have more overall impact on the final results.

## 2. RELATED WORK

### 2.1 Triple Scoring

There are varied motivations for the task of determining the relevance score for a given triple based on a type-like relation. This score is a crucial component of any ranking system for entity queries, because it allows for the sorting (and subsequent retrieval and presentation) of results based on their relevance. A significant research has been conducted by Bast, Buchhold and Haussmann [1] who propose a variety of algorithms to compute relevance scores for knowledge base triples from type-like relations. These algorithms combine methods of word counting that analyze text about a person from the English Wikipedia, as well as probabilistic models that utilize the maximum likelihood estimation (e.g., Generative Model). Their top performing algorithm achieved 80% accuracy using an unsupervised learning approach applied to a text corpus.

---

[1] http://broccoli.cs.uni-freiburg.de/wsdm-cup-2017

## 2.2 Word Embedding Representation

Vector based word representation has a successful history of use in information retrieval and computational semantics as an implementation of the long-standing linguistic hypothesis that words that occur in similar contexts tend to have similar meanings [5]. More recently, Mikolov et al. [8] introduced his work known as *Word2vec*, which was shown to be able to capture also linguistic regularities of both semantic ('king' to 'man' is like 'queen' to 'woman') and syntactic nature ('ran' to 'run' is like 'laughed' to 'laugh'). It has two main formulations for the target prediction: skip-gram and continuous bag-of-words (CBOW). We test both models and no significant differences were observed. For this reason and also because it is a more computationally efficient model for larger datasets, following Mikolov et al. [7], we employ the continuous bag-of-words (CBOW).

The CBOW model learns to predict the word in the middle of a symmetric window based on the sum of the vector representations of the words in the window. Computation of the output word probability distributions by the standard softmax classifier is expensive ($O(|W|)$), since this requires summing all of the words in the vocabulary, which is generally very large. As an alternative, negative sampling estimates the probability of an output word by learning to distinguish it from draws from a noise distribution. The number of these draws (number of negative samples) is given by a parameter $k$. In addition, very frequent words are not very informative as context features. Word2vec implements a method to minimize their effect (and simultaneously improve performance). More precisely, words in the training data are discarded with a probability that is proportional to their frequency: $p(discard|w) = 1 - \sqrt{\frac{\rho}{f(w)}}$ where $f(w)$ is the frequency of the word $w$ and $\rho$ is a parameter that controls how aggressive the sub-sampling is.

## 3. APPROACH

The given training set provided a limited number of samples to learn from (both the profession and nationality training files contain approximately 670 tuples). We assumed that this information alone would not produce a good prediction score, as we were requested to predict the relevance score for approximately 340,000 people. In order to boost our predictive model performance, our approach consists of two main steps: 1) expanding the current knowledge based on an entity's immediate environment, and more specifically, its nearest neighbors, and 2) using the knowledge gained in order to predict the relevance score for a given tuple.

**Figure 1: The flow of the proposed solution**

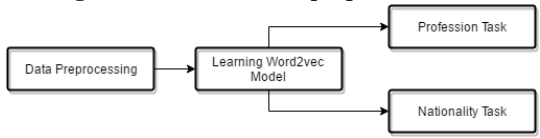

Figure 1 describes our approach as follows. It starts from the preprocessing performed on the data (described in section 3.1). Thereafter, we build a word2vec model by inputting the text (after preprocessing) into the model as described in section 3.2. Finally, we describe how each relation task, profession and nationality, utilizes the word2vec model to learn and predict the relevance score as described in sections 3.3 and 3.4, respectively.

### 3.1 Data Preprocessing

The `wiki-sentences` file provided by WSDM Cup 2017 contains more than 33 million sentences from Wikipedia. Each sentence contains an annotation for at least one person (e.g., "[Barack_Obama|Obama]", "[Bob_Dylan|Bob Dylan]"), where the term on the left of the "|" is the name of the person as a single word, and the term on the right represents the word as it appears in the original sentence. Each person from the `persons` file is associated with some number of wiki-sentences (ranging from 68,662 sentences for the most frequently mentioned person, to three sentences for the least frequently mentioned person in this file).

We conducted several preprocessing steps to train the word2vec model (described in section 2.2). First, we removed sentences that contain only one word as they are uninformative, since the model tries to learn from the context of the words in each sentence. We also replaced the annotated expression (e.g., "[Barack_Obama|Obama]") for each name with a single word term (e.g., Barack_Obama). By doing so, each name appears uniformly in the `wiki-sentences` file.

The model treats each word individually; thus, we've adjusted professions or nationalities that are more than one word long, replacing the space between words with the character '_' (for example, "American football player" was converted to "American_football_player").

Using pre-trained vectors[2], which were trained on part of the Google News dataset (about 100 billion words), we created a static list of nationality mapping in which each country is mapped to the relevant nationality. For example, we ran this query on the existing vectors and found that X = Canadian: 'United_states_of_america' to 'Canada' is like 'American' to 'X'? Then, we added the nationality (based on the mapping) to each sentence in which a county name appears. This should help the model better differentiate the context of nationality when a country name appears; for example, if "United States of America" appears in a sentence in the context of a person, it most likely means that the person is an "American".

Finally, we removed stopwords such as "on", as well as punctuation characters such as '+'.

### 3.2 Learning Word2vec Model

We trained our model with the word2vec gensim package[3]. The package implements both the skip-gram and CBOW approaches [8]. We use the CBOW model in order to learn word embedding and apply the negative sampling method. In addition, sub-sampling is applied as described in section 2.2. After data preprocessing (described in section 3.1), the resulting text is tokenized using the Stanford tokenizer[4], and each word is converted to lowercase. The final dataset contains approximately 29 million valid sentences. We set the dimension of the word embeddings to 300; the number of negative samples per positive sample is set at 15; $\rho$ is set at 1e-4; $C$ is set at 5; and the number of iterations is 50. It is important to note that we experimented with different settings of the various parameters, and no significant change in the results was observed.

In each relation of the two-pronged task, we used the *most_similar* method implemented in the gensim package to identify the Top-N most similar words. This method computes cosine similarity (referred to as "similarity score" in this paper) between a simple mean of the projected weight vectors of the given words and the vectors for each word in the model.

### 3.3 Profession Relation Task

In the profession relation aspect of the task we were asked to predict a relevance score for a given tuple <person,profession>. Our approach consists of two main steps: 1) expanding the current knowledge based on an entity's immediate environment, referred to as "Learning Step" (section 3.3.1), and 2) using the knowledge gained in order to predict the relevance score for a given tuple, referred to as "Prediction Step" (section 3.3.3).

---
[2]https://code.google.com/archive/p/word2vec/
[3]https://radimrehurek.com/gensim/models/word2vec.html
[4]http://nlp.stanford.edu/software/tokenizer.shtml

### 3.3.1 Learning Step

The main purpose of the learning step is to discover as much about a person as possible given the current knowledge available. It consists of two main stages:

1. **Score Propagation** - In the propagation stage, we aimed to identify the professions associated with each person from their nearest neighbors, and as the number of iterations increased, we found that we were able to identify new people that can be informative about their environment.
2. **Score Normalization** - In the normalization stage, we aimed at aggregating a list of scores, thereby establishing a normalized score. People with higher similarity scores have greater impact on the person currently being considered and thus become dominant in the calculation of the final score.

The learning algorithm (Algorithm 1) begins by initializing the *overall_state* dictionary with the first 70% of the tuples from the training set, together with their relevance scores (line 1). Thereafter, we set the values of *n_after* and *n_before* to MAX_INTEGER and 0, respectively (lines 2-3). Both variables denote the number of distinct people before and after the propagation stage. Before the propagation starts, we count the number of distinct people (line 5). Then, the ScorePropagation() subroutine (described in Algorithm 2) produces a dictionary, which contains an entry for each person discovered in the propagation stage, with their professions and the associated relevance scores (line 6). Thereafter, we merge the entries from both the *overall_state* and *new_state* dictionaries (line 7) and count the number of distinct people in the *overall_state* (line 8).

**Algorithm 1** Profession Learning Algorithm
1: $overall\_state \leftarrow train\_set$
2: $n\_after \leftarrow MAX\_INTEGER$
3: $n\_before \leftarrow 0$
4: **while** $(n\_after - n\_before) > 0$ **do**
5:   $n\_before \leftarrow distinct\_person(overall\_state)$
6:   $new\_state \leftarrow \text{SCOREPROPAGATION}(overall\_state)$
7:   $overall\_state \leftarrow new\_state \cup overall\_state$
8:   $n\_after \leftarrow distinct\_person(overall\_state)$
9: **end while**

The learning step ends when the difference between the number of people before and after the propagation stage is 0 (line 4), or in other words, when we weren't able to find any new people in the previous iteration, and there is no point in continuing this process. To improve efficiency, the output of the training step was saved to an external file which will be used in the prediction step which is described in section 3.3.3.

The ScorePropagation() subroutine is a key function of the profession learning algorithm (Algorithm 1). As shown in Algorithm 2, the input to the function is *overall_state* dictionary which represents the knowledge we have acquired so far. It begins by initializing the *current_state* dictionary to an empty set (line 2). Thereafter, the score of each person in the *overall_state* dictionary is propagated to his/her Top-10 similar people. To reduce noise, we only considered pairs whose similarity score is higher than 0.4 (lines 3-11). Once the score propagation loop has been completed, the algorithm returns a normalized score for each tuple <person,profession> in the *current_state* dictionary (line 12).

**Algorithm 2** Score Propagation Algorithm
1: **function** SCOREPROPAGATION(overall_state)
2:   $current\_state \leftarrow \emptyset$
3:   **foreach** $person \in overall\_state$ **do**
4:     $x \leftarrow person\_profession\_list$
5:     $top\_similar \leftarrow model.most\_similar(person, topn = 10)$
6:     **foreach** $similar\_person \in top\_similar$ **do**
7:       **if** $similarity\_score \geq 0.4$ **then**
8:         $current\_state[similar\_person] \leftarrow (x, similarity\_score)$
9:       **end if**
10:    **end for**
11:  **end for**
12:  **return** SCORENORMALIZATION($current\_state$)
13: **end function**

**Score Normalization**: Given a set of tuples: $T_{i,j} = \{(rel\_score, sim\_score) \mid person = i, profession = j\}$ where $rel\_score$ is the relevance score and $sim\_score$ is the similarity score, the score normalization can be expressed as follows:

$$Norm\_Score_{ij} = \frac{1}{Z_{ij}} \sum_{k \in T_{i,j}} k.rel\_score \times k.sim\_score \quad (1)$$

where:

$$Z_{ij} = \sum_{k \in T_{i,j}} k.sim\_score$$

$i \in \{1, .., M\}$ indexes the relevant person and $j \in \{1, .., N\}$ indexes the relevant profession of each tuple, respectively.

The first iteration starts with the knowledge in the training set as our initial state. In the second iteration, the number of distinct people in the *overall_state* will be larger, because of the new people identified in the previous iteration, and so on. In each iteration we reveal up to $10 \times N$ new people, where $N$ is the number of people in the *overall_state*, and 10 is the number of similar people in the propagation stage. The strength of this process is demonstrated best in cases in which each of the 10 most similar people for a given new person is also new – in these cases the knowledge obtained is multiplied 10 times! This process enabled us to uncover 264,874 people after just 10 iterations, a figure which represents over 77% of the total number of distinct people listed in the `persons` file. In the propagation stage, we aimed to identify the professions associated with each person from their nearest neighbors, and as the number of iterations increased, we found that we were able to identify new people that can be informative about their environment. However, from the eleventh iteration we were unable to identify new undiscovered people. Section 3.3.3 provides an explanation of how the people that were not discovered during the propagation stage are handled.

### 3.3.2 Method Illustration

In this section, we demonstrate the learning step on real data from the training set, including the score propagation and normalization stages. For simplicity, we assume that Barack Obama is the only person in our training set together with his relevance scores which are presented in Table 1. The first step is to initialize the *overall_state* dictionary with Barack Obama's records.

| Profession | Score |
|---|---|
| Politician | 7 |
| Laywer | 0 |
| Law Professor | 1 |
| Author | 0 |

**Table 1: Relevance scores for Barack Obama**

In the next step, referred to as the ScorePropagation() subroutine, we were able to find Obama's Top-10 similar people by using our model's *most_similar* method (described in section 3.2). The "Score" column in Table 2 below represents the cosine similarity (similarity score) between Barack Obama and his Top-10 similar people. The score value is in the range of [-1,1]; the higher the score is, the more likely the two people are similar:

|  | Name | Score |
|---|---|---|
| 1 | george_w._bush | 0.868378 |
| 2 | bill_clinton | 0.842833 |
| 3 | hillary_clinton | 0.824706 |
| 4 | john_mccain | 0.791295 |
| 5 | joe_biden | 0.770765 |
| 6 | john_kerry | 0.761532 |
| 7 | mitt_romney | 0.754377 |
| 8 | george_h._w._bush | 0.754377 |
| 9 | jimmy_carter | 0.746474 |
| 10 | ronald_reagan | 0.731227 |

**Table 2: Obama's Top-10 similar people**

Obama's four profession scores will be propagated, with respect to the similarity score, to each of the people on the above list of similar people. In this case, this results in a new entry in the *current_state* dictionary for george_w._bush as follows: {"Politician": (7, 0.868378), "Lawyer": (0, 0.868378), "Law professor": (1, 0.868378), "Author": (0, 0.868378)}. Each profession is associated with a tuple, where the first element represents the relevance score and the second element is the similarity score. This process is performed for the rest of Obama's similar people.

In this way, a personal dictionary will be formed for each person, which contains the professions he/she has accumulated, along with the scores obtained in the propagation stage. If george_w._bush receives a profession from a new person that is already part of his dictionary, this additional knowledge is added to his existing tuple list. For example, if a new person close to george_w._bush (with a similarity score of 0.67656) has a relevance score of 6 for "Politician", this information is propagated to Bush whose updated tuple list will appear as follows, with a new tuple appended to "Politician": {"Politician": (7, 0.868378), **(6, 0.67656)**, "Lawyer": (0, 0.868378), "Law professor": (1, 0.868378), "Author": (0, 0.868378)}. Just before the ScorePropagation() subroutine ends, the ScoreNormalization() subroutine in which the scores are normalized is initiated. Following the previous example for george_w._bush the calculation for "Politician" is:

$$Z_{ij} = (0.868378 + 0.67656) = 1.544$$

$$Norm\_Score_{ij} = \frac{1}{1.544}(0.868378 \times 7 + 0.67656 \times 6) = 6.562$$

where $i$ corresponds to the index of 'george_w._bush' in the set of persons, and $j$ to the index of 'politician' in the set of professions. Note that the person with the highest similarity score (0.868378) has the greatest impact on the final score. The final result in this case will be 7 since this is the closest integer.

### 3.3.3 Prediction Step

Given a tuple $<i,j>$ where $i \in persons$ and $j \in professions$, we perform the following steps:

1. If tuple $<i,j>$ exists in the output file from the learning step, the score is returned as is.

2. Otherwise, we find the Top-10 similar people to person $i$, and we try to determine if profession $j$ exists in the profession lists of the Top-10 similar people. Because person $i$ can obtain relevance scores from each of the Top-10 similar people, the final list may contain several scores. Then, we normalize the score with respect to the similarity score in the same manner described in section 3.3.1.

3. If the steps above do not produce any relevance scores, we utilize the word2vec model to find the most similar professions to profession $j$ (we look at the Top-5 similar professions – see Table 3). The intuition behind this is that related professions (e.g., Singer and Singer-songwriter) are likely to be represented as nearby vectors in a vector space (explained in section 2.2), and they can be indicative regarding whether a particular person "belongs" to a profession $j$.

|  | Name | Score |
|---|---|---|
| 1 | singer-songwriter | 0.835767 |
| 2 | musician | 0.808166 |
| 3 | vocalist | 0.802527 |
| 4 | songwriter | 0.733304 |
| 5 | performer | 0.649689 |

**Table 3: Examples of the Top-5 similar professions for singer based on the word2vec model**

Then, for each similar profession $k$, we check the similarity score between person $i$ to profession $k$. Similarity scores that are greater or equal to the defined threshold (=0.4) received a value of 1. Finally, we total the results and obtain a score ranging from 0 to 7 (mimicking the actions of the judges which will be described later in section 4); in the step described here, we utilize similar professions to assess how the tuple $<i,j>$ is related.

## 3.4 Nationality Relation Task

In the nationality relation aspect of the task we were asked to provide a relevance score for a given tuple <person, nationality>. We were unable to take the same approach used for the profession task, because our basic assumption that similar people share the same nationality is not correct in this case. For example, "Barack Obama" can be close to "Vladimir Putin" in our latent space, because they have semantic relation (both are politicians), however they obviously don't share the same nationality. In the next section we explain how we overcome this issue.

### 3.4.1 Learning Step

Our basic aim is to determine how a specific person is related to any of the 100 countries listed in the `nationalities` file. To assess the strength of the relation, we counted the number of occurrences of each country in the person's Wikipedia entry. We noticed that in many Wikipedia entries the person is associated with a nationality, as well as a country name; for example, in the Wikipedia entry for "Alanis Morissette", she is listed as being Canadian-American, and the entry also mentions the United States of America and Canada. Therefore, instead of only considering the country name, we also looked for the nationality mapped to the country based on the nationality mapping described in section 3.1.

Initially, for each entity in the persons file we used the corresponding Wikipedia page (implemented with Wikipedia API for python[5]). We also manipulated the text, removing stopwords and punctuation characters, and ignoring digits. For each entity (e.g., "Alanis Morissette"), we counted the number of occurrences of each country/nationality. In the final step these numbers were normalized by the maximum number of occurrences observed until this point. The normalized result is a number between 0 and 1, where the entry (country in this case) with the maximum number of occurrences receives a score of 1, and the nationalities without any occurrence obtain a score of 0. As the final relevance score is a number between 0 and 7, we multiply the normalized value by 7 and round this number to the closest integer. To increase efficiency the output of the learning step was saved to an external file which will be used in the prediction step (described in section 3.4.2).

Note that we were unable to find a Wikipedia entry for some entities; this may be related to technical issues (e.g., DisambiguationError if the page is a disambiguation page, or a PageError if the page doesn't exist). In next section we describe how we address this situation.

### 3.4.2 Prediction Step

The prediction phase varies, depending on whether a Wikipedia entry exists for the person or not. The case in which there is a Wikipedia entry is relatively simple, because the value obtained in the learning step is used as the prediction score. In the second case, in which we are unable to find a Wikipedia entry for a person, we used the word2vec model. For each of the 100 nationalities, we calculated the similarity score based on our model. Those nationalities with a negative score were discarded, because this provides a good indication that they are not semantically related. Finally, all of the values were normalized based on the maximum similarity score observed to ensure that the highest score is assigned to the most similar nationality. Each score was multiplied by the maximum relevance score that we can give to each tuple (7) to ensure that our final prediction value will range from 0 to 7.

## 4. EVALUATION RESULTS

Evaluation is based on three relevance measures, two score-based and one rank-based:

1. Accuracy (ACC) - The percentage of triples where the score is within ±2 of ground truth.
2. Average Score Difference (ASD) - The absolute score difference, averaged over all triples in the test set.
3. Kendall's Tau (TAU) - The similarity of rankings for same subject (percentage of pairs with transpositions), averaged over all subjects.

Each of the training files contains scores for tuples (<person,profession> or <person,nationality>) which were judged by seven human judges. Each judge was requested to determine whether each tuple is primarily relevant (= 1) or secondarily relevant (= 0), and their decisions were added together so that the maximum score each tuple could obtain was 7.

The WSDM Cup 2017 challenge does not provide a test set to evaluate our method; in the absence of a test set, we must rely upon the training files which contain hundreds of pairs. Note that for the profession relation, we utilized 70% of the training set for the learning step (described in section 3.3.1), leaving just 30% of the profession data for validation and testing. Our official submission achieved the following results: ACC - 0.74, ASD - 1.74, and TAU - 0.35. However, since the accuracy measure counts any score within ±2 from the ground truth as an accurate prediction, there is no point in assigning scores of 0, 1, 6, or 7. After the official submission, we applied the 2-5 truncation, so that a score of 0 or 1 changes to 2, and a score of 6 or 7 changes to 5. Our method achieved better results after making this adjustment: ACC - 0.81, ASD - 1.74, and TAU - 0.36. Our ranking in the competition improved accordingly, and our submission moved from 8th place to 4th place. Of course there is a trade-off when applying the 2-5-truncation: ACC goes up , but it usually makes the ASD and TAU considerably worse (which means that the value increases, larger values are worse for ASD and TAU). Additional information regarding the final results can be found in the overview paper [2].

## 5. CONCLUSION

In this paper, we proposed a comprehensive and flexible method to compute relevance scores for each triple relation, a method that can be used to work effectively in both cases of the two-pronged task: profession and nationality. We demonstrated the significant potential of using neural embedding techniques to learn the relevance scores from type-like relations, achieving nearly 74% accuracy.

The results of the current work indicate important directions for future work. A future direction worth further investigation is the impact of using a combined approach based on the word2vec model and the co-occurrence technique explored in the nationality relation. A combined model that can take into account weighting, in which each model has a different level of importance in the final score (e.g., $w_1 \times Model_1 + w_2 \times Model_2$), could be considered as well.

## References

[1] H. Bast, B. Buchhold, and E. Haussmann. Relevance scores for triples from type-like relations. In *SIGIR*, pages 243–252. ACM, 2015.
[2] H. Bast, B. Buchhold, and E. Haussmann. Overview of the Triple Scoring Task at the WSDM Cup 2017. In *WSDM Cup*, 2017.
[3] K. Bollacker, C. Evans, P. Paritosh, T. Sturge, and J. Taylor. Freebase: a collaboratively created graph database for structuring human knowledge. In *Proceedings of the 2008 ACM SIGMOD international conference on Management of data*, pages 1247–1250. ACM, 2008.
[4] T. Cover and P. Hart. Nearest neighbor pattern classification. *IEEE transactions on information theory*, 13(1):21–27, 1967.
[5] Z. S. Harris. Distributional structure. *Word*, 10(2-3):146–162, 1954.
[6] S. Heindorf, M. Potthast, H. Bast, B. Buchhold, and E. Haussmann. WSDM Cup 2017: Vandalism Detection and Triple Scoring. In *WSDM*. ACM, 2017.
[7] T. Mikolov, Q. V. Le, and I. Sutskever. Exploiting similarities among languages for machine translation. *arXiv preprint arXiv:1309.4168*, 2013.
[8] T. Mikolov, I. Sutskever, K. Chen, G. S. Corrado, and J. Dean. Distributed representations of words and phrases and their compositionality. In *Advances in neural information processing systems*, pages 3111–3119, 2013.
[9] F. Ricci, L. Rokach, and B. Shapira. *Introduction to recommender systems handbook*. Springer, 2011.

---

[5] https://pypi.python.org/pypi/wikipedia/